\documentclass[epjCONF,columns]{svjour} 
\usepackage{graphics}
\usepackage[varg]{txfonts} 
\usepackage[latin1]{inputenc}
\session-title{Hadron Collider Physics Symposium 2011}
\begin{document}
%
\title{A Measurement of the ATLAS Di-Muon Trigger Efficiency in Proton-Proton Collision at $\sqrt{s}=7$ TeV}
\author{ATTILIO PICAZIO\thanks{\email{attilio.picazio@cern.ch}} on behalf of the ATLAS Collaboration}
\institute{D\'epartement de physique nucl\'eaire et corpusculaire, Universit\'e de Gen\`eve, 1211 Geneva 4, Switzerland}
\abstract{The B physics programme of the ATLAS experiment includes measurements of production cross sections, searches for rare $B$-decay signatures which are sensitive to new physics at the TeV energy scale and studies of CP violation effects in B-events, such as $B_{s}^{0}\rightarrow J/\psi \phi$ and $B_{d}^{0}\rightarrow J/\psi K_{s}^{0}$. The key to the detection of these B signals in ATLAS is to achieve a high trigger efficiency for low-$p_{T}$ di-muon events, whilst keeping an acceptable trigger rate. ATLAS developed two separate approaches for triggering on di-muon events from resonances such as $J/\psi$ and Upsilon ($\Upsilon$). The first approach is to start from a di-muon trigger selected at Level-1 while the second is based on dedicated Level-2 algorithms. The performance for these triggers has been studied using collision data at $\sqrt{s}=7$ TeV collected in 2011.
} 
\maketitle
\section{Introduction}
\label{intro}
\begin{figure}
\begin{center}
\resizebox{1.\columnwidth}{!}{%
 \includegraphics{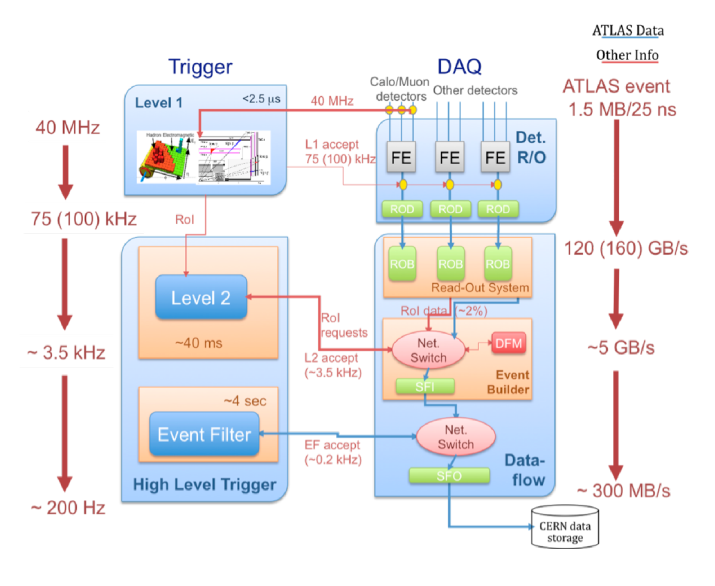} }
\caption{A schematic of the ATLAS Trigger Systems and its interfaces to the detector and readout systems.}
\label{fig:ATLASTrig}
\end{center}
\end{figure}

The ATLAS detector \cite{RefJ} is a multi-purpose apparatus with a cylindrical geometry covering almost the entire solid angle around the interaction point. Closest to the beam pipe are silicon based tracking detectors and straw-tube transition radiation detectors, located inside a superconducting solenoid that provides a 2T magnetic field. Outside the solenoid, fine-granularity LAr electromagnetic (EM) calorimeters with an accordion geometry provide coverage up to $|\eta| < 3.2$. An iron-scintillating hadronic calorimeter extends to $|\eta| < 1.7$ while copper and LAr technology is used in the endcap region covering the region $1.7 < |\eta| < 3.2$. In the very forward region, $3.2 < |\eta| < 4.5$, copper and tungsten LAr calorimeters provide measurements of the EM and hadronic showers. The Muon Spectrometer (MS) consists of three superconducting toroidal magnets with precision tracking provided by Monitored Drift Tubes (MDT) up to $|\eta| < 2.4$ and Cathode Strip Chambers (CSC) in the very forward region. Triggering for muons is provided using Resistive Plate Chambers (RPC) in barrel region and Thin Gap Chambers (TGC) in the forward region up to $|\eta| < 2.4$. The ATLAS detectors provide input to the three levels of the trigger system that provides fast filtering capabilities and capture events of interest to physics with high efficiency. The ATLAS Trigger System is designed to reduce the output storage rate to ~200 Hz (about 300 MB/sec) from an initial LHC bunch crossing rate of 40 MHz. The first level trigger, Level-1 (L1), is hardware based. The L1 receives its inputs from the Calorimeter, Muon and additional forward detectors and provides a decision with a latency of less than 2.5 $\mu s$. So-called Regions of Interest (RoIs) are defined around the L1 signatures; the L2 algorithms (which also access inner tracking information) operate only in these regions, reducing the time taken for the L2 decision to be made. The L2 and third-level Event Filter (EF) together comprise the High Level Trigger (HLT), and these are based on software algorithms which run on a large computing farm close to the detector. A schematic diagram of the ATLAS trigger is shown in Fig. \ref{fig:ATLASTrig}.

The LHC started single-beam operations in 2008 and achieved its first proton-proton (pp) collisions in 2009. In 2010, the LHC began colliding protons on protons at a centre of mass energy of 7 TeV. During a prolonged operation in 2010, ATLAS collected 45 $pb^{-1}$ of data with luminosities ranging from $10^{27}$ $cm^{-2}s^{-1}$ to $2 .10^{32}$ $cm^{-2}s^{-1}$. At the time of this conference, ATLAS had collected 4.8 $fb^{-1}$ in 2011, with instantaneous luminosities exceeding $3.65\times10^{33}$ $cm^{-2}s^{-1}$.

ATLAS prepared a programme for measurements of production cross sections both of B-hadrons and Onia in central proton- proton collisions. Dedicated triggers based on muon, di-muon or electron signatures were designed to accommodate the large statistics provided by LHC.  With high statistics collected in 2010 and 2011 B-hadron properties and CP violation (via the exclusive decay $B_{s}\rightarrow J/\psi\phi$) are being studied.
Crucial to all of these measurements is the performance of the muon triggers and the offline muon reconstruction and identification, as well as the Inner Detector (ID) performance.

\section{ATLAS Di-Muon Trigger for B-Physics}
\label{sec:1}
\begin{figure}
\resizebox{1.\columnwidth}{!}{%
 \includegraphics{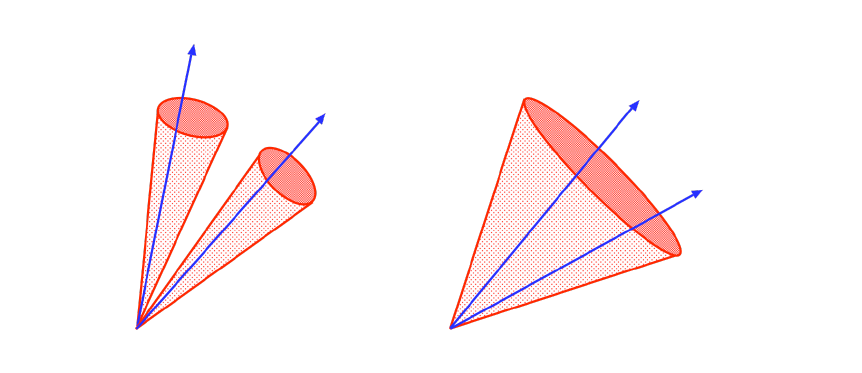} }
\caption{A schematic picture di-muon trigger, topological trigger (left) and seeded by a single L1 muon (right).}
\label{fig:1}       
\end{figure}

The output rate of the L1 trigger at a luminosity of $10^{33}$ $cm^{-2}s^{-1}$ is expected to contain about 20 kHz of events where one muon exceeded the transverse momentum ($p_{T}$) threshold of 6 GeV/c. In 2010 it has been possible to include even lower $p_{T}$ thresholds, down to the lowest threshold achievable in the Level 1 hardware. At L2 this rate of events must be reduced to 1-2 kHz, of which 5-10 \% are available for channels of interest only to B-physics. Currently this goal is achieved for L2 muon triggers by first confirming that a muon over the nominal threshold is reconstructed in the MS, and then confirming that there is a matching track in the ID. This selection criterion removes many muons from K and $\pi$ decays, but does not by itself produce the required rate reduction. To achieve the required rate $p_{T}$ thresholds need to be raised and many interesting $B$-events are likely to be filtered out. To achieve high efficiency for these signals at L2, specific di-muon trigger algorithms were developed. Since most of the relevant processes involve a di-muon (either directly from the B-hadron in the case of rare decays, or via a $J/\psi$, $\psi$(2s), or an $\Upsilon$) a significant proportion of the heavy flavour and quarkonia decays falling in the detector acceptance can be captured by ATLAS.
There are two possible approaches at L2 for selecting di-muon events from a resonance such as $J/\phi$ and $\Upsilon$. The first is to start from a selected di-muon trigger at L1. In this approach each muon is confirmed separately, and the invariant mass of the pair can then be calculated and a mass window cut applied. This trigger will be referred to as the ``topological di-muon trigger''.
An alternative approach is to start with a L1 single muon trigger and search for the second muon track inside a wide $\eta-\phi$ region of the ID and then extrapolating the track to the MS to tag muon tracks. Since this method does not explicitly require the second muon at L1, it has an advantage for reconstructing $J/\psi$ even if the second muon has a $p_{T}$ lower than the lowest L1 threshold. This trigger will be referred to as the ``TrigDiMuon'. The basic ideas for this two approaches (topological and TrigDiMuon) are illustrated in Fig. \ref{fig:1}.

The performance of the topological di-muon trigger approach was studied using 2011 data at $\sqrt{s}=7$ TeV. A precise understanding of the trigger efficiency is essential for a wide range of analyses, especially cross section measurements. A data-driven method of estimating the di-muon trigger efficiency, based on Bayes Theorem, has been developed. It is independent of the angular distribution of the signal muons from the J/$\psi$ or $\Upsilon$, and can be applied on any data stream without any bias on the efficiency measurement, even if the data stream has been selected by a muon trigger item.
In this approach the efficiency for a di-muon trigger item can be written as:
\begin{equation}
\epsilon(T_{2\mu})=\epsilon(T_{1\mu})\bigotimes\epsilon(T_{2\mu}|T_{1\mu})\bigotimes\frac{1}{\epsilon(T_{1\mu}|T_{2\mu})}
\end{equation}
where $\epsilon(T_{2\mu})$ is the di-muon trigger item efficiency, $\epsilon(T_{1\mu})$ is the efficiency for a single muon trigger item (for example the EF decision with a $p_{T}$ threshold of 4 GeV/c), $\epsilon(T_{2\mu}|T_{1\mu})$ is the conditional probability for the di-muon trigger given the single muon trigger and $\epsilon(T_{1\mu}|T_{2\mu})$ represents the conditional probability for single muon trigger given the di-muon trigger. The latter term ($\epsilon(T_{1\mu}|T_{2\mu})$) is equal to unity for topological trigger and close to unity for the TrigDiMuon trigger.
The $\epsilon(T_{1\mu})$ term has been obtained with the Tag and Probe method using 2011 data with the same procedure described in \cite{RefT}. The conditional probabilities can be measured directly from data - in particular, the efficiency $\epsilon(T_{2\mu}|T_{1\mu})$ corresponds to the number of events for which both $T_{2\mu}$ and $T_{1\mu}$ fired, divided by the total number of events that passed the $T_{1\mu}$ trigger threshold. This approach for evaluating $\epsilon(T_{2\mu})$ was tested on a large Monte Carlo sample and on 2010 collision data showing very good agreement between data and Monte Carlo simulation \cite{RefP}. In this work the results obtained on 2011 collision data are shown.  

\section{Di-Muon Trigger Efficiency on 2011 Data}
\label{sec:2}
\begin{figure} 
\begin{center}

\resizebox{0.75\columnwidth}{!}{%
 \includegraphics{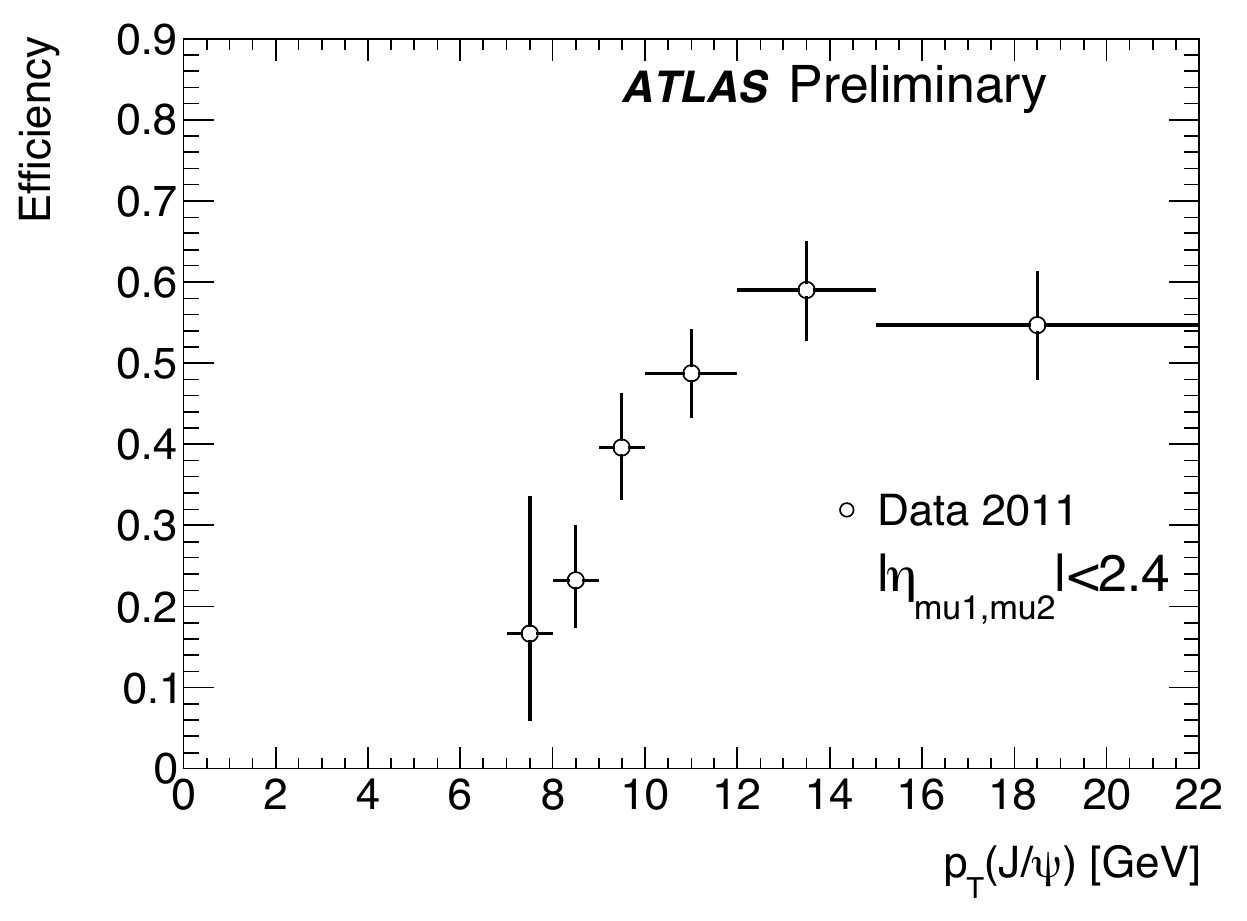} }
\caption{Efficiency of the di-muon trigger EF\_2mu4\_Jpsimumu with respect to $J/\psi$ $p_{T}$, evaluated on candidate $J/\psi\rightarrow\mu^{+}\mu^{-}$ events with respect to offline combined muon reconstruction.}
\label{fig:Jpsipt}
\end{center}
\end{figure}

\begin{figure}
\begin{center}
\resizebox{0.75\columnwidth}{!}{%
 \includegraphics{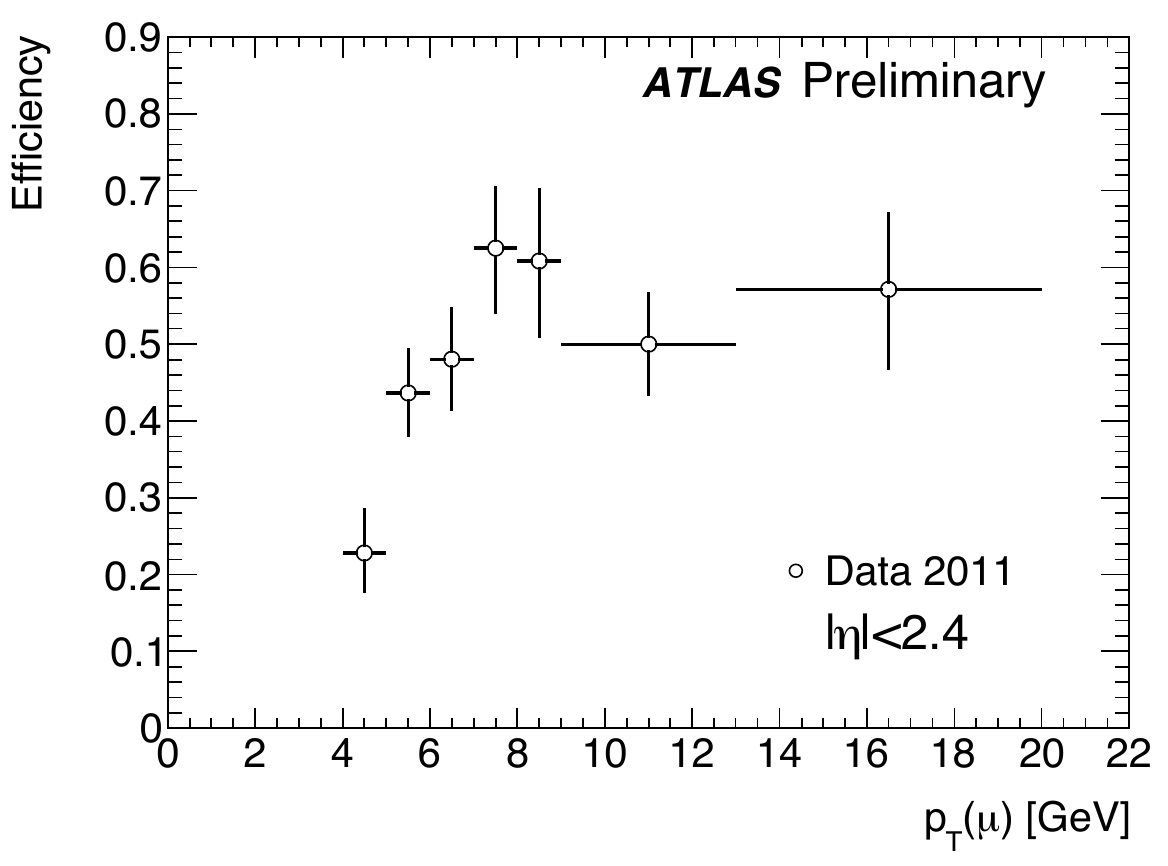} }
\caption{Efficiency of the di-muon trigger item EF\_2mu4\_Jpsimumu with respect to muon $p_{T}$ (the higher $p_{T}$ muon) evaluated on candidate $J/\psi\rightarrow\mu^{+}\mu^{-}$ events with respect to offline combined muon reconstruction.}
\label{fig:dimupt}
\end{center}
\end{figure}

\begin{figure}
\begin{center}
\resizebox{0.75\columnwidth}{!}{%
 \includegraphics{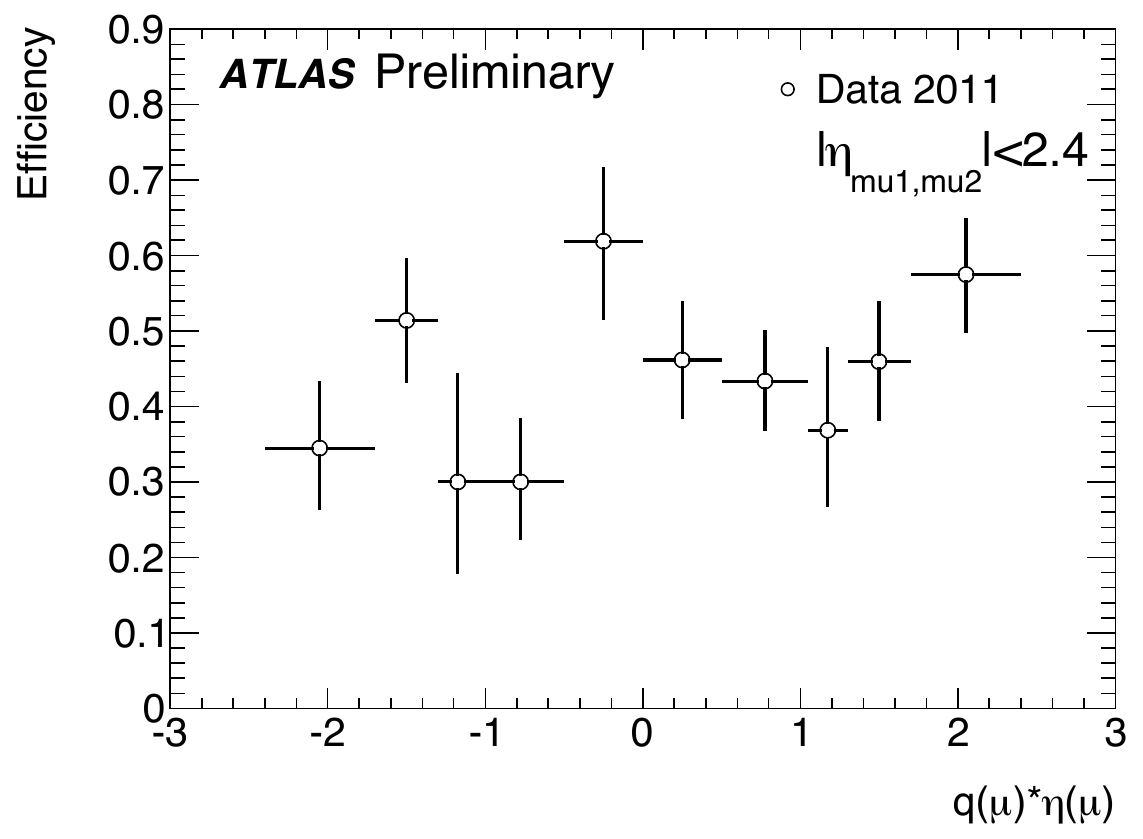} }
\caption{Efficiency of the di-muon trigger EF\_2mu4\_Jpsimumu with respect to muon $\eta \ast q$ (the higher $p_{T}$ muon) evaluated on candidate $J/\psi\rightarrow\mu^{+}\mu^{-}$ events with respect to offline combined muon reconstruction.}
\label{fig:dimueta}
\end{center}
\end{figure}

\begin{figure}
\begin{center}
\resizebox{0.75\columnwidth}{!}{%
 \includegraphics{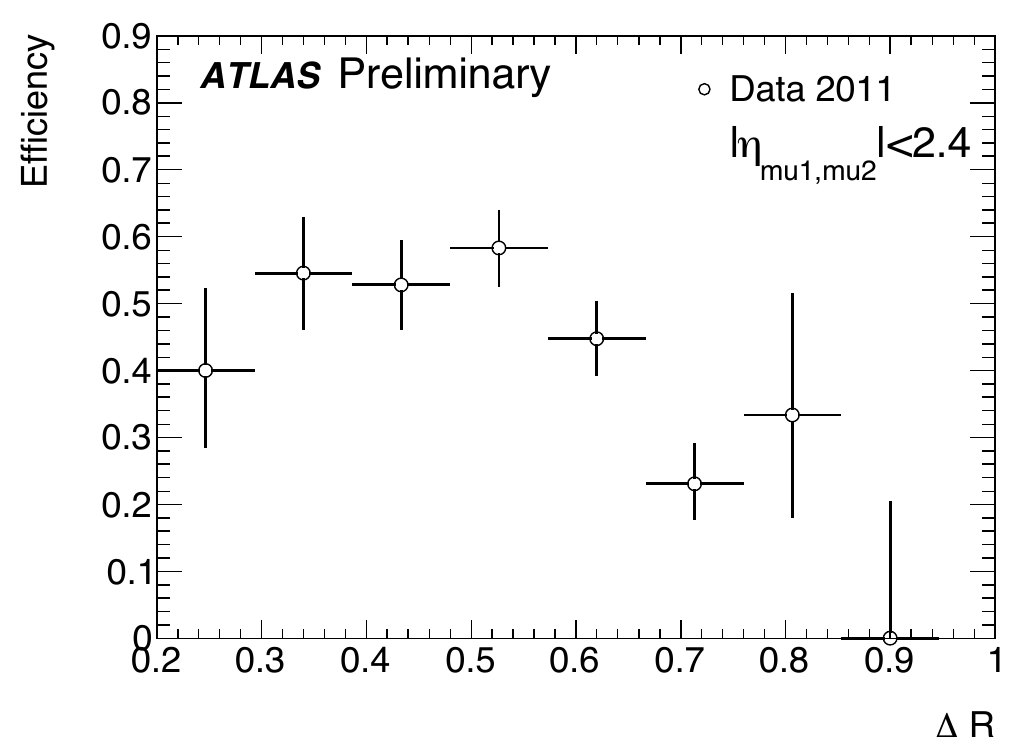} }
\caption{Efficiency of the di-muon trigger EF\_2mu4\_Jpsimumu with respect to the $\Delta R$ (angular separation) of the two muons evaluated on candidate $J/\psi\rightarrow\mu^{+}\mu^{-}$ events with respect to offline combined muon reconstruction.}
\label{fig:dimuDR}
\end{center}
\end{figure}

The 2011 data sample used to measure the di-muon trigger efficiency for the trigger item EF\_2mu4\_Jpsimumu corresponds to an integrated luminosity of 1.25 $fb^{-1}$. This topological trigger item operates in the invariant mass range $[2.8 - 4.5]$ GeV/$c^{2}$ where the larger contribution to the di-muon final state comes from the $J/\psi\rightarrow\mu^{+}\mu^{-}$ decays and,  as already mentioned, it requires that both muons pass the EF\_mu4 trigger. The method described previously was deployed for this trigger item, using the single trigger the EF decision with a $p_{T}$ threshold of 4 GeV/c.  In Fig. \ref{fig:Jpsipt} 
the efficiency of the di-muon trigger EF\_2mu4\_Jpsimumu with respect to $J/\psi$ $p_{T}$ is shown, evaluated on candidate $J/\psi\rightarrow\mu^{+}\mu^{-}$ events with respect to offline combined muon reconstruction. From this plot the trigger efficiency for the physics signature selected by this item can be understood. The statistical uncertainty is dominated by the large prescale factor applied on the EF\_mu4 trigger item. In principle, the di-muon trigger efficiency depends on $p_{T}$, $\eta$ and $\phi$ of both muons inside the di-muon pair and on the opening angle between the two muons ($\Delta R$), defined as:
\begin{equation}
\Delta R=\sqrt{\Delta \phi^{2}+\Delta \eta^{2}}
\end{equation}
In Fig. \ref{fig:dimupt} the efficiency of the trigger EF\_2mu4\_Jpsimumu with respect to the $p_{T}$ of the higher $p_{T}$ muon in the $J/\psi\rightarrow\mu^{+}\mu^{-}$ candidate pair is shown.
The efficiency with respect to the product of $\eta$ and the charge of the higher $p_{T}$ muon is shown in Fig. \ref{fig:dimueta}. The latter variable is useful to observe the effect on the efficiency due to the toroidal magnetic field in the forward region of the detector that affect mainly low-$p_{T}$ muons.    
Finally, in Fig. \ref{fig:dimuDR} the trigger efficiency is reported with respect to the $\Delta R$ of the muons. The drop at large $\Delta R$ is a kinematical effect due to the small mass of the J/$\psi$. 

\section{Conclusions}
\label{sec:3}
A method based on Bayes theorem has been developed to measure the di-muon trigger efficiency both for topological and TrigDiMuon items. It has been applied to measure the trigger efficiency of the topological di-muon trigger on the 2011 data sample collected by the ATLAS detector at LHC at $\sqrt{s}= 7$ TeV.
%


\begin{thebibliography}{}
\bibitem{RefJ}
The ATLAS Collaboration, \textit{The ATLAS experiment at the CERN Large
Hadron Collider}, JINST 3 (2008) S08003
\bibitem{RefA}
The ATLAS Collaboration, \textit{Expected Performance of the ATLAS Experiment}, arXiv:0901.0512 CERN-OPEN-2008-020
\bibitem{RefT}
The ATLAS Collaboration, \textit{A measurement of the ATLAS muon reconstruction and trigger efficiency using J/$\psi$ decays}, ATLAS- CONF-2011-021, http://cdsweb.cern.ch/record/1336750/
\bibitem{RefP}
The ATLAS Collaboration, \textit{A measurement of the ATLAS di-muon trigger efficiency in proton-proton collisions at $\sqrt{s}= 7$ TeV}, ATL-DAQ-PROC-2011-044, http://cdsweb.cern.ch/record/1401937



\end{thebibliography}
\end{document}